# Stochastic Carbon Footprint Tracing Methods in Power Systems

Jiashuo Hu, *Student Member, IEEE*, Xiao-Ping Zhang, *Fellow, IEEE, Fellow, CSEE,* Youwei Jia, *Member, IEEE*

*Abstract*—**As the penetration of distributed energy resources (DER) and renewable energy sources (RES) increases, carbon footprint tracking requires more granular analysis results. Existing carbon footprint tracking methods focus on deterministic steady-state analysis where the high uncertainties of RES cannot be considered. Considering the deficiency of the existing deterministic method, this paper proposes two stochastic carbon footprint tracking methods to cope with the impact of RES uncertainty on load-side carbon footprint tracing. The first method introduces probabilistic analysis in the framework of carbon emissions flow (CEF) to provide a global reference for the spatial characteristic of the power system component carbon intensity distribution. Considering that the CEF network expands with the increasing penetration of DERs, the second method can effectively improve the computational efficiency over the first method while ensuring the computational accuracy on the large power systems. These proposed models are tested and compared in a synthetic 1004-bus test system in the case study to demonstrate the performance of the two proposed methods.**

## I. INTRODUCTION

AS a major contributor to anthropogenic climate change, carbon emissions are receiving increasing attention [1]. Various countries and regions, including China, the UK, and the EU have set targets for carbon neutrality. The power system, contributing to about 40% of fossil fuel emissions, is a crucial target for emission reduction, which is thus essential to an accurate carbon emission measure method [2]. In this process, the establishment of an effective methodology for the observation, measurement, and analysis of carbon emissions by component is the basis for a fair and efficient implementation of incentives for the power system [3]-[10].

In the realm of carbon emissions research within power systems, considerable attention has been directed towards measurements on the generation side, predominantly utilizing life cycle analysis (LCA) methods [11]-[13]. LCA comprehensively scrutinizes carbon emissions across the entire lifecycle of electricity production. However, an equally critical aspect lies in addressing emissions on the load side, a pivotal element in the decarbonization process of power systems; and a profound understanding of carbon footprint per unit of energy consumed at the load side is of paramount importance [9]. To map the carbon footprint trajectory from generation to load, the carbon emission flow (CEF) methodology was introduced [10]. This approach focuses on translocating the carbon footprint to the consumption side, based on the concept of a "virtual" carbon emission flow, a notion increasingly recognized in the domain of international energy trading [14], [15].

The CEF model is identified as a more precise tool for distinctly pinpointing the carbon footprint of load-side components in power systems. Reference [16] demonstrated the application of CEF in a national-scale power system to achieve the 2020 emission reduction target. Additionally, references [17] and [18] have integrated CEF methodologies with carbon optimization problems, serving as indicators for computing their carbon optimization outcomes. These studies serve as pivotal benchmarks for calculating outcomes in emission optimization scenarios. Recent studies also amalgamate the CEF approach with electricity and carbon markets, establishing a trading mechanism predicated on load carbon footprint [7], [19]. Most studies deploy the CEF method to tackle specific load-side bus carbon emission responsibility issues, such as the optimization of carbon footprints of particular components over a defined period.

On the other hand, RES and DER penetration in the power system illustrated an increasing trend [20] [21]. Due to the highly stochastic and unpredictable characteristics of RES, the carbon emissions responsibility of the power system under relatively high-RES penetration for load-side components cannot be represented by studies of specific time sections. The traditional CEF analysis results based on specific time section power flow analysis cannot provide a long-term reference for power systems with high new energy penetration rates. Another consequence of the high unpredictability of RES is that traditional generator needs to adjust their output more frequently to maintain the stability of the power system. As noted in references [22] and [23], the generation efficiency of conventional generating units varies with its output fluctuations during its electrical producing process, which affects its real-time carbon emission intensity (CEI). This variation has an impact on the actual carbon emissions of the generating unit during the dispatch process of the power system. Currently, most studies of CEF use static average parameters of the generating unit as input parameters for

This work was supported in part by the National Natural Science Foundation of China (72371123), and Shenzhen Sustainable Development Research Program (KCXST20221021111210023). (*Corresponding author: Youwei Jia*)

J. Hu, Y. Jia are with the Dept. of Electronic and Electrical Engineering, Southern University of Science and Technology, 518088, Shenzhen, China, and J. Hu is also with Department of Electrical, Electronic and Systems Engineering,

School of Engineering, University of Birmingham, Birmingham B15 2TT, UK. (e-mails: jxh1480@student.bham.ac.uk, jiayw@sustech.edu.cn)

X.-P. Zhang is with the Department of Electronic, Electrical and Systems Engineering, School of Engineering, University of Birmingham, Birmingham B15 2TT, UK. (e-mails: x.p.zhang@bham.ac.uk)

average emission factor (AEF) [15-20]. Under the high penetration of RES, the output of the generator fluctuates more frequently; thus, the AEF method may underestimate the environmental impact of the power plant [26]. And power systems with high DER penetration rates need to expand carbon emission analysis to larger-scale distribution networks which significantly increases the computational burden of the CEF analysis method. Generally, high penetration of RESs and DERs in the power system imposes higher requirements on the calculation efficiency of carbon footprint analysis methods.

To establish a decarbonized power system, it is crucial to adopt a methodology that traces the load-side carbon footprint from a system-level global perspective. However, as far as we know, existing carbon tracing methods are based on deterministic steady-state analysis, which cannot reflect the impact of RES uncertainty on the carbon footprint of load-side components. In response to this situation, this paper presents two stochastic models designed specifically for tracing carbon footprint. The proposed model could consider the impact of components with uncertainty characteristics (e.g., RES, electric vehicles (EVs), etc.) on the carbon footprint tracing of the power system to provide a global reference for probabilistic analyses. The main contributions of this paper are summarized below:

- Stochastic Carbon Emissions Tracing Method: These innovative models offer a robust probabilistic analysis tool within the CEF methodology framework, which could ascertain the statistical distribution of the load-side carbon footprint, providing a clear representation of its probability distribution under high uncertainties.

- Ultra-efficient stochastic carbon footprint tracing method: Under this stochastic carbon footprint tracing framework, the paper proposes a "virtual bus" concept, which aggregates load-side components with identical emission intensities to significantly enhance computational efficiency under the stochastic analysis as ultra-efficient stochastic carbon footprint tracing method.

- Spatial characteristics of the power system carbon intensity: This paper established statistical methods to provide the probability distribution of system components' carbon footprint under the impact of power system topology. The analysis results are considered to a statistical reference for developing carbon reduction incentives and hence constructing a carbon market for the power system under the framework of CEF methodology.

The proposed model was tested on a synthetic 1004-bus test system under the case study. The analysis results show that the proposed stochastic carbon footprint tracing model could reveal the complex non-linear relationship between the system's RES penetration and the load carbon footprint under the extensive range power system topology. The ultra-efficient stochastic carbon footprint tracing model proposed based on this method can well cope with the problem of reduced efficiency of probabilistic computation brought by large-scale networks. The remainder of this paper is organized as follows. Section II describes the framework of the proposed stochastic carbon emissions tracing model. Section III introduces the concept of the virtual bus in the first methodological framework introduced in Section II to significantly improve the computational efficiency and an ultra-efficient carbon footprint tracing method is proposed. Section IV provides results from a case study using a combined model by the modified IEEE 16-bus transmission system combined with the IEEE 33-bus distribution test system. Finally Section VI concludes this paper.

## II. STOCHASTIC CARBON FOOTPRINT TRACING METHOD

### A. Carbon Emission Responsibility Allocate Based on CEF

The CEF model defines carbon emission from the production of electrical energy as one of the attributes of electrical energy and assumes that this attribute is distributed with the power flow based on the proportional sharing principle [11]. There are three main variables in the CEF model, including the CEF rate, branch carbon intensity and bus carbon intensity. All these variables with the unit of $kgCO_2/kWh$, the value of these variables reflects the carbon emission responsibility of the component per unit of electricity consumed. The CEF model is built on the bus carbon intensity, which calculates the carbon responsibility of the bus by obtaining a weighted average of the injected power flow, which could be formulated below as (1) and (2) mathematically. $\Omega_i^G$ is the set of local generators injecting power into the bus $i$ ; $\Omega_i^{B^+}$ denote the set of buses of upstream component of the bus $i$ ; $P_{ij,t}$ is the active power flow between bus $i$ and $j$ the direction of power flow determined by the value of $\ell_{ij,t}$, which denotes the branch carbon intensity; $P_{G_i}$ and $e_{G_i}$ denote the active power output of generator located at bus $i$ and generator's carbon intensity; $e_{i,t}$ is the bus carbon intensity of bus. All these variables are calculated within time $t$.

$$e_{i,t} = \frac{\sum_{i \in \Omega_i^G} P_{G_i} \cdot e_{G_i} + \sum_{i \in \Omega_i^{B^+}} |P_{ij,t}| \cdot \ell_{ij,t}}{\sum_{i \in \Omega_i^G} P_{G_i} + \sum_{i \in \Omega_i^{B^+}} |P_{ij,t}|} \tag{1}$$

$$\ell_{ij,t} = \begin{cases} 0, & if \ P_{ij,t} > 0 \\ e_{j,t}, & if \ P_{ij,t} > 0 \end{cases} \tag{2}$$

### B. Modeling of RES

The RES in this paper is represented by the wind farm (WF). The output of WF depends on the wind speed, and the Weibull distribution is commonly used to describe its probabilistic property. The PDF and CDF of the Weibull distribution are shown in (4) and (5). $x$ represents the wind speed; $\lambda$ and $k$ are parameters to determine the distribution of wind speed.

$$f(x) = \begin{cases} \frac{k}{\lambda} (\frac{x}{\lambda})^{k-1} e^{-(x/\lambda)^k} & x \geq 0 \\ 0 & x < 0 \end{cases} \tag{4}$$

$$F(x) = \begin{cases} 1 - e^{-(x/\lambda)^k} & x \geq 0 \\ 0 & x < 0 \end{cases} \tag{5}$$

The wind speeds $x$ that follow the above distribution could be obtained by means of sampling. The relationship between the wind speed and output power of WF could be described in (6).

$P_{wind}$ denote the output power of WF and $P_{rate}$ is the rate output of WF, there are three key parameters of WF, $v_{in}$, $v_{cut}$, and $v_{rate}$ where are the cut-in, cut-out, and rate wind speed of WF respectively.

$$
\begin{cases}
P_{wind} = 0 & x < v_{in} \\
P_{wind} = P_{rate} \cdot \dfrac{x - v_{in}}{v_{rate} - v_{out}} & v_{in} < x < v_{rate} \\
P_{wind} = P_{rate} & v_{rate} < x < v_{out} \\
P_{wind} = 0 & x > v_{out}
\end{cases}
\tag{6}
$$

### C. Modeling of Load

The actual power under the power system could be assumed to be a random variable and follow a certain probability. There are a range of studies available to inform this and could be plugged into this stochastic carbon emissions tracing model [28]-[32]. In this paper, two main types of loads are considered, i.e., the bus base load and EV charging loads respectively.

The normal distribution is used frequently to simulate bus base load [28]-[31]. Therefore, the normal distribution in this paper to determine the load value at different sampling scenarios follows in (7), (8), and (9).

$$
L_t \sim N(\mu_t, \sigma_t^2)
\tag{7}
$$

$$
\varphi(x) = \frac{1}{\sigma_t \sqrt{2\pi}} e^{-\frac{(x - \mu_t)^2}{2\sigma_t^2}}
\tag{8}
$$

$$
\phi(x) = \int_{-\infty}^{x} \varphi(u) du
\tag{9}
$$

$L_t$ is the random variable of the bus base load at typical sampling scenarios. $\mu_t$ and $\sigma_t^2$ are the mean and variance of $L_t$; $\varphi(x)$ and $\phi(x)$ are the probability density and cumulative density function of $L_t$.

For EV charging load, this paper uses the plug-in hybrid EV (PHEV) charging demand model in reference [24] to the model of the overall charging demand for PHEVs. The EV demand of a single EV charging station was determined via fit of the Weibull distribution by the test of Kolmogorov-Smirnov goodness-of-fit test (K-S test) with significance level $\alpha = 5\%$; the parameters are provided by reference [24].

### D. Marginal Carbon Emissions Model of Traditional Generator

After the RES and load side have been determined for each sampling scenario, the traditional generator takes on the role of offsetting the gap between the generation and consumption side. Under different sampling scenarios, the output of the traditional generator is no longer stable at a particular value and fluctuations in its output will have a variable effect on its power generation efficiency and its CEF rate. The CEF value of the traditional generator is affected by many factors. In this paper, we have extended the linear model in [22] to consider the effect of RES uncertainty by considering the scenario where the traditional generator output fluctuates around its optimum output. The model simplifies the relationship between output power and CEF rate of the generator as a segmented linear function is shown in (10). $a_i^{down}$, $b_i^{down}$, $a_i^{over}$, and $b_i^{over}$ are the generator parameters, which are determined by the types and size of the generator. $P_{Gi}$ is the output of the generator, $P_{Grate}$ and $P_{Glim}$ is the designed optimal and maximum output power of the generator. $r_i$ is the CEI corresponding to the output value of the generator.

$$
r_i = \begin{cases}
a_i^{down} - b_i^{down} P_{Gi} & 0 < P_{Gi} < P_{Grate} \\
a_i^{over} + b_i^{over} P_{Gi} & P_{Grate} < P_{Gi} < P_{Glim}
\end{cases}
\tag{10}
$$

Under each sampling scenario, the output of RES and value of load are determined following the probability distribution setting first; the output of the traditional generator is then determined to fill the difference between the load side and the RES generation. After determining the output value of the conventional generator, the stochastic carbon emission tracing model can determine the CEI value of the conventional generator according to (10), which is used as an input to the calculation of the system-wide carbon footprint.

### E. Scenarios Sampling of Stochastic Carbon Emissions Tracing Model

Monte Carlo Simulation (MCS) is an effective tool for qualifying various uncertainties in power systems. In the framework of MCS, the uncertainty of the input parameter is represented by its corresponding probability distribution. The output analysis result be presented in the form of probability distributions or probability density functions, such as the voltage, power, and losses, etc. Under the stochastic carbon emissions tracing model, after obtaining a definitive carbon emission responsibility result from CEF result; the wide range of load-side carbon emission responsibility possible scenarios could be generated through extensive random sampling based on the known probability distributions of loads, traditional generator output, and renewable energy outputs from [20], [18], [24], and [25]. The scenarios sampling method under the stochastic carbon emissions tracing model needs to be calculated as follows:

$$
S = f(H)
\tag{3}
$$

The $S$ is the carbon emissions responsible for each component of the power system as the output result; $H$ is the input vector, including the load with electrical vehicle, the output of the RES, and the output of traditional generation units, etc.; which are represented as stochastic variables in conventional power system $\{h_{i+1}, h_{i+1}, ..., h_n\}$. And the $f(H)$ is the CEF model with the inputs and outputs of probability distributions. Based on the MCS methods, the carbon intensity probability distributions of power system components can be accurately represented under uncertainty in the RES.

## III. Ultra-efficient Stochastic Carbon Footprint Tracing Method

### A. Concept of Virtual Bus under CEF Analysis

Based on the power flow direction between the components of the power system, the interrelationships could be represented as upstream and downstream. As the example shown in Fig.1; the component that injects the power to the bus $b_i$ is called the upstream component of the bus ($B_i^+$), which include the transmission line $Path_{i-1,i}$ and $Path_{i-2,i}$ with the connected buses ($b_{i-1}$ and $b_{i-2}$), and the generator $G_1$. The bus $b_i$ as a mixer of power from the $B_i^+$, the carbon intensity of $b_i$ determined by $B_i^+$ based on the CEF model as shown in (1) and (2).

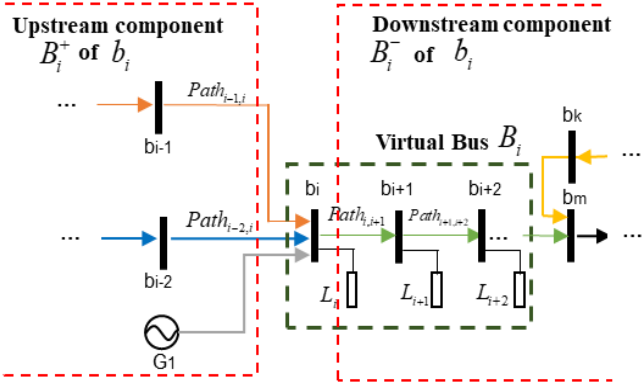

Fig. 1. Diagram of the Virtual Bus Concept.

Based on the CEF model, the downstream component $b_i$ includes the load $L_i$, transmission line $Path_{i,i+1}$, bus $b_{i+1}$, and other component receive power from $b_i$, be represented as $B_i^-$ in Fig 1 with same carbon intensity due to there being no other sources. This carbon intensity translation process be interrupted at $b_m$ due to $b_m$ has another upstream component thus changing the proportion of power in the components downstream of $b_m$. Under CEF model, the value of carbon intensity of the component only receives power from $b_i$ are same, which could be aggregated into a component under CEF calculations. This paper refers to this aggregated component as a virtual bus as $B_i$ in Fig 1.

According to the definition of a virtual bus, the bus would be the starting bus of a virtual bus if connected to local DERs. After identifying the virtual bus's starting bus, follow the power flow direction, iterating through downstream buses until reaching the start of another virtual bus. This process is shown in Fig. 3. All components within the range of this virtual bus are considered as a single node in the CEF model.

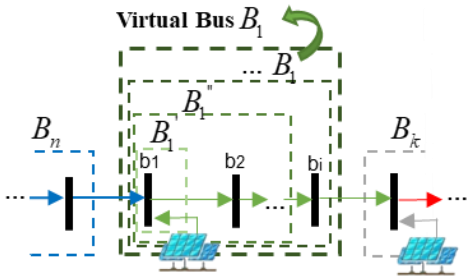

Fig. 2. Iteration of Virtual Bus Distribution in Network.

### B. Decomposition of Network Topology with Virtual Buses

Due to the large amount of system components, further decentralization of the carbon footprints under the CEF model to the network globally and the implementation of probabilistic analyses under stochastic carbon footprint tracking require significant computational resources. However, the concept of a virtual bus could significantly simplify the topology of radial topology networks thus reducing the requirement of carbon footprint tracing under total system level. In the system topology, the locations of DERs in the distribution network are identified. Based on the unidirectional and radial topology of the distribution network flow, the range from nodes of the DERs connected node to downstream node until another local DER is aggregated into a virtual bus. According to equations (1) (2), all components within a virtual bus have identical characteristics in terms of their carbon footprint in the CEF model calculations. Thus, simplifying the power system based on the virtual bus method does not affect the accuracy of CEF calculations.

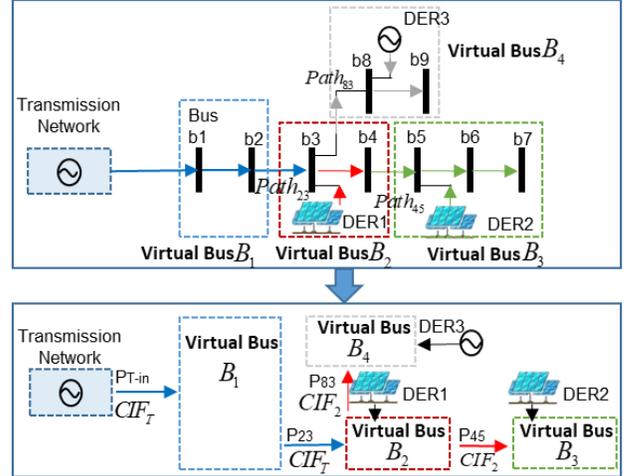

Fig. 3. Simplified Topology with Virtual Buses.

For instance, in a typical 9-node distributed network as represented in Fig 3, DERs located at nodes 3, 5, and 8 decompose the network into 4 virtual nodes. In the CEF calculations, this 9-node distribution network can be regarded as a connection of 4 virtual nodes, thereby reducing the number of nodes under the CEF calculations. In the stochastic carbon footprint tracking model, the topological network can be effectively simplified by introducing a virtual bus approach, thus improving the computational efficiency during the calculation of the load-side component's carbon footprint. Generally, the power system topology would be simplified to virtual bus connections. This simplified network would then be utilized in the stochastic carbon footprint tracing model to analyze the probabilistic distribution of carbon footprints of different components in the system. The framework of ultra-efficient stochastic carbon footprint tracing model as shown in Fig. 4.

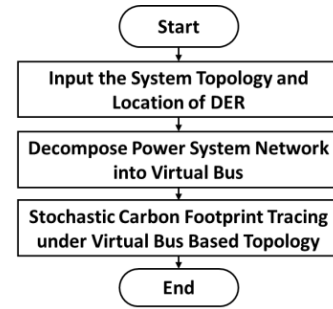

Fig. 4. Ultra-efficient Stochastic Carbon Footprint Tracing Framework.

## IV. Case Study

To demonstrate the viability and efficiency of the stochastic carbon emissions tracing model, the case study will be demonstrated on a synthetic 1004-bus test system using the workstation Inter(R) Core (TM) i7-12700H. The test system is a combination of a modified IEEE-16 bus transmission system and the modified IEEE-33 bus distribution systems which are connected to the transmission system bus. For the generation side, the test model uses the balancing bus as a coal-based generator, and there are two combined cycle gas-based generators connected at bus 2 and bus 3; two WFs connected at bus 6 and bus8 as the system's RES input. In addition, a total of 30 small PV units are connected to the distribution network as DERs at random locations, and the DER capacity is 20% of the local load. The case study will include five scenarios with different RES penetration rates (0%, 20%, 40%, 60%, and 80%, respectively). For the load side, the based bus load of all buses followed the setting in [31] and the EV load is considered as part of the consists load of the distribution network on IEEE16-bus transmission system buses 9, 10, and 11 following the setting in [24].

The case study consists of three parts: the first part illustrates the improvement of model computational efficiency by the virtual bus method; the second part quantifies the proportion of the responsibility for the generator's carbon emissions that should be taken by the load-side components in different scenarios; finally, the third section explores the variation in load-side components carbon footprint under different RES penetration rates.

TABLE I Comparison of Computational Time Between Virtual Bus Topology vs Original System

| Sampling Number | 1000 | 5000 | $10^4$ | $2 \times 10^4$ |
|---|---|---|---|---|
| Virtual-Bus Topology (s) | 0.14 | 0.68 | 1.44 | 2.85 |
| Primary Topology (s) | 23.85 | 127.89 | 295.74 | 588.18 |

### A. Efficiency Increase via Virtual Bus Method

To verify the efficiency of the virtual bus method in calculating the load-side carbon footprint distribution of the power system, we compared the model calculation time of the carbon footprint distribution using the simplified power system topology of the virtual bus method with the original power system topology for different samples. Under the CEF model framework, the simplified power system topology nodes have a total of 76 virtual buses, i.e., the size of the node matrix to be processed by the CEF model in each sampling is reduced from 1004 to 76 nodes while guaranteeing the computational accuracy, which significantly reduces the total elapsed time of the Monte Carlo sampling method under the stochastic carbon footprint tracing model. The computational time of the stochastic carbon tracing model under primary topology and virtual-bus topology is shown in Table I. A comparison of the times calculated by the procedure shows that the virtual-bus-based power system could significantly reduce the time required for the carbon footprint analysis, which significantly improves the efficiency of the calculation; the efficiency gain increases progressively with the number of samples taken. Using the virtual bus method for load-side carbon footprint analysis of the power system will save a significant amount of computational resources.

### B. Allocation of Responsibility for Carbon Emissions from Generator

To investigate the changes in the total carbon emissions of the generator and the proportional share of emissions responsibility of each component on the load side under increased RES penetration, the stochastic carbon emissions tracing model was used to analyze a coal-based generator (G1), which is located at bus1. Fig 5 illustrates the total carbon emission charging under different RES penetrations. With the RES penetration increase, the carbon emission of the coal-based generator illustrated the decrease trend.

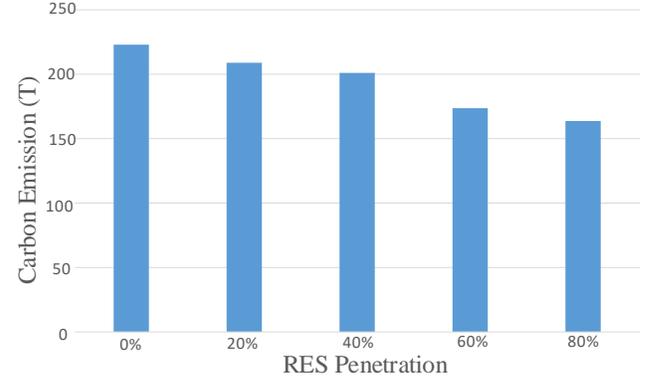

Fig 5. Carbon emission of G1 under different RES penetrations.

The proposed stochastic carbon emission tracing model could allocate the carbon emission responsibility to the load side. Taking the scenarios with 40% and 80% RES penetration of each load side component that should be responsible for the G1 carbon emission, as shown in Fig 6. With the RES penetration increase, the emission of coal-based generators illustrated a decreasing trend, however, the load-side emission responsibility did not show a linear charge. Compared with the responsibility of load side component under 40% and 80% RES penetration, the carbon emission responsibility of loss and load at Bus 3, 4, and 13 show a decrease trend; but the load carbon footprint at Bus 6, 9, 10, 11, 14 show an increase trend with other load side component keep roughly same.

This analysis result provides a new perspective on the participation of generators in the carbon market under the CEF framework. Currently, the cost of trading carbon allowances for generating units is generally spread evenly across the cost of each unit of electricity produced. As a result, downstream load-side components will receive an equal share of the economic impact via the same tariff change for the electricity. This case demonstrates to some extent that it is not fair to spread the carbon market transaction costs of a generating unit equally to each individual customer under the CEF analysis framework, as each customer should not be equally responsible for carbon emissions. The responsibility of carbon emission that the load side should take depends on the topology of the power system and other relevant factors, which is highly non-linear with the system RES penetration increase.

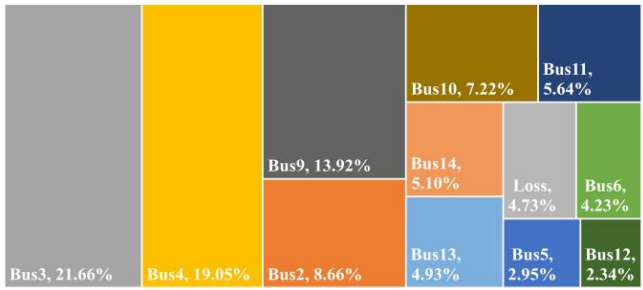

(a). Carbon footprint under 40% RES penetration

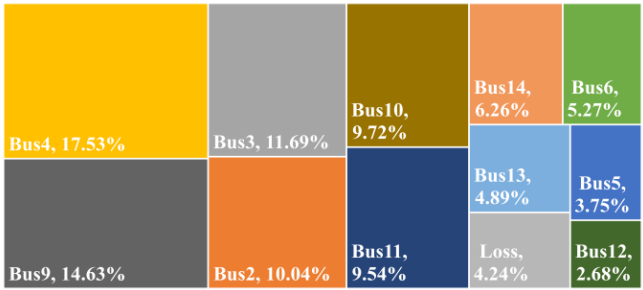

(b). Carbon footprint under 80% RES penetration

Fig 6. Responsibility of load side component on G1 carbon emission.

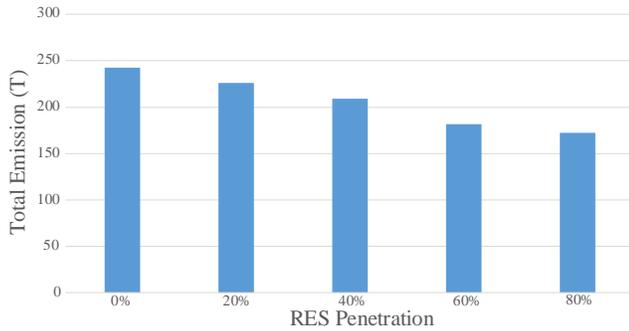

Fig 7. The total emission of systems under different scenarios.

### C. Effect of Different RES Penetration on Carbon Footprint of Load-side Components

For the carbon emissions observed from the generator side, the total emission of the power system shows a significant negative correlation with the RES penetration of the system, as shown in Fig 7. This downward trend means that the increase in the RES penetration rate of the system will bring about a reduction in the overall carbon emissions of the power system, which is consistent with the general.

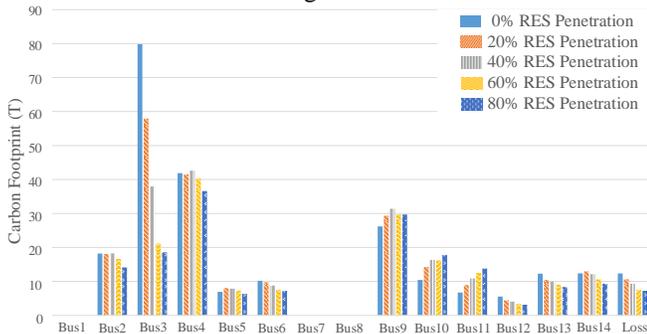

Fig 8. Carbon emission is caused by load-side components under different scenarios.

The next step is to determine the net emission of load-side under different RES penetrations with the CEF analysis framework. For the load-side component, the load at Bus 2, Bus 6, and Bus 9 (connected to local gas-based, connected to local RES, net load, respectively) and line losses the examples to illustrate the effect of RES penetration rate on the load side carbon footprint. For the carbon emissions caused by the load-side components, the emission under different scenarios is shown in Fig 8. Bus 7 and 8 are the voltage balance nodes with no active load and net emission.

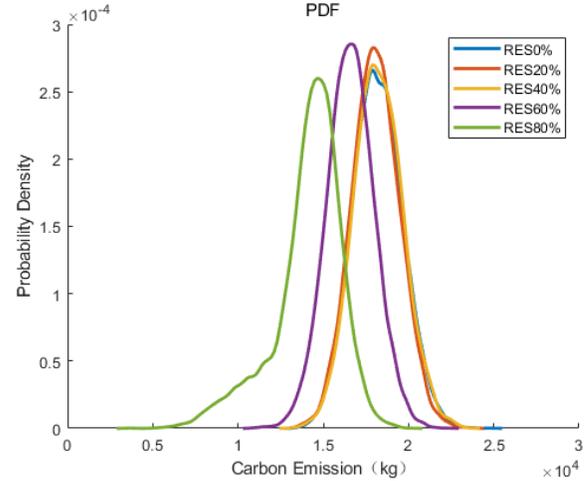

Fig 9. Carbon footprint probability density of load at Bus2.

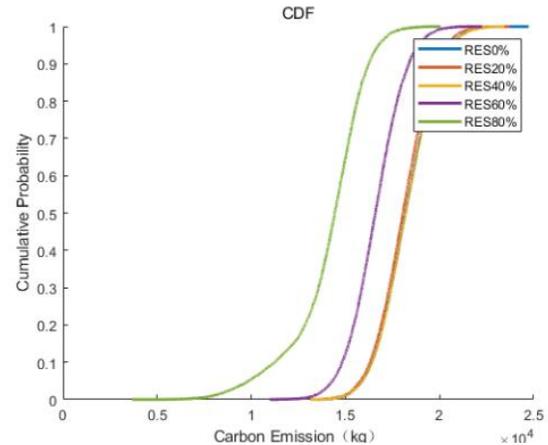

Fig 10. Carbon footprint cumulative density of load at Bus2.

With the RES penetration increase, the total emissions caused by the load at Bus 2 keep almost the same value under the first three scenarios. It should be noted that the carbon footprint of load located at Bus 2 under 40% system RES penetration is roughly higher than the value under 20% system RES penetration, showing the opposite trend of common sense. The reason is the uncertainty of the RES can cause conventional generator output to deviate from its maximum fuel efficiency output rating to maintain system frequency. The statical analysis results of the probability density distribution and cumulative distribution of the carbon emissions caused by the load at Bus 2 are shown in Fig 9 and 10 also illustrate this trend.

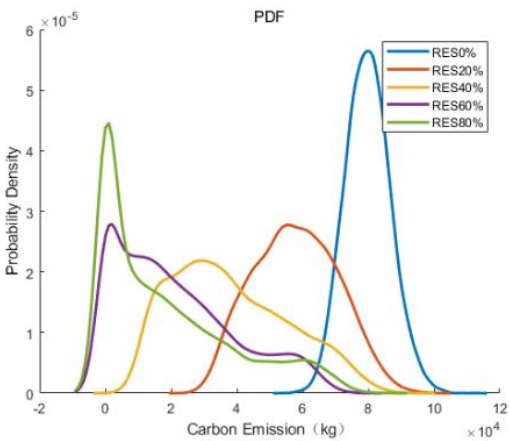

Fig 11. Carbon footprint probability density of load at Bus 6.

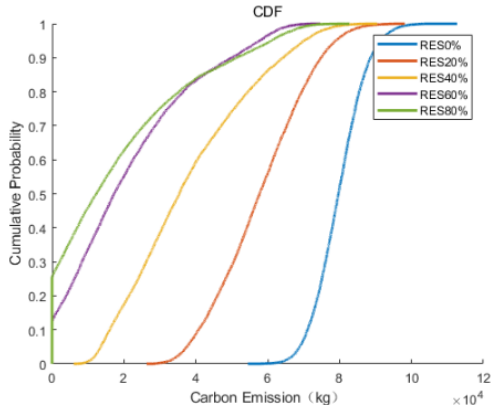

Fig 12. Carbon footprint cumulative density of load at Bus 6.

For the load located at Bus 6 which connects to the RES, the load carbon footprint at Bus 6 has a relatively low value under all the Scenarios. With the RES penetration increase, the emission caused a decreasing trend; the statical analysis result of the probability density distribution and cumulative distribution of the carbon emissions caused by the load at Bus 6 is shown in Fig 11 and 12.

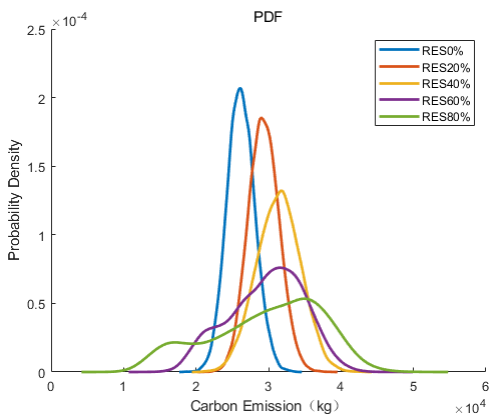

Fig 13. Carbon footprint probability density of load at Bus9.

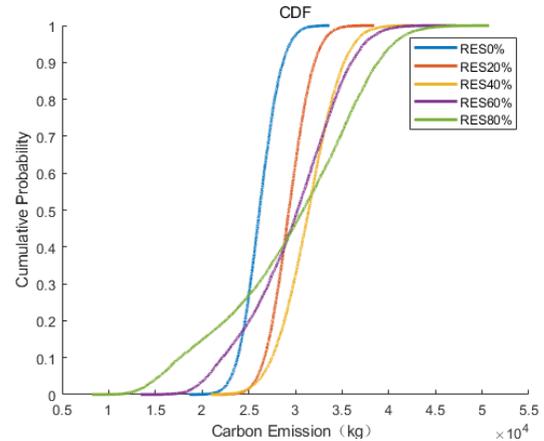

Fig 14. Carbon footprint cumulative density of load at Bus9.

The situation changes on Bus 9, 10, and 11, which can be observed in Fig 6. For the loads located on Bus 9, 10, and 11, the load carbon footprint does not decrease with the increase in RES penetration. The probability and cumulative density distribution for the load on Bus 9, for example, are shown in Fig 13 and 14.

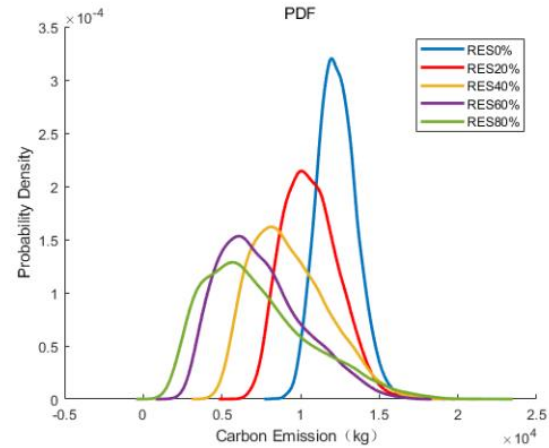

Fig 15. Carbon footprint probability density of transmission losses.

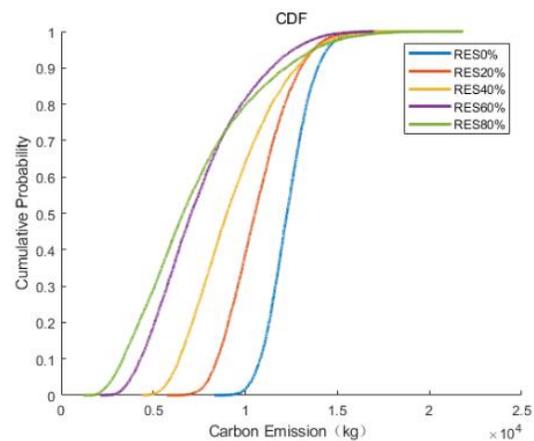

Fig 16. Carbon footprint cumulative density of transmission losses.

For the carbon footprint of system losses, with the RES penetration increase, the loss continues to decrease. The probability and cumulative density distribution for the losses are shown in Fig 15 and 16.

The result of the analysis presented in this case shows that although the total carbon emissions of the system decrease as the RES penetration increases, this trend does not apply to the load-side carbon emissions responsibility analysis. As illustrated in Fig 17, affected by the topological structure of the power system, the average carbon density of different node energies is affected by changes in the total system penetration rate, which is related to the power system, network topology generator location, etc. The net carbon emissions of the system caused by load-side components have a highly non-linear relationship with the total system RES penetration. While increasing the RES penetration reduces the total emissions of the system, the different buses would be affected differently, and it is unfair to homogenize their emissions responsibility change under the CEF analysis framework.

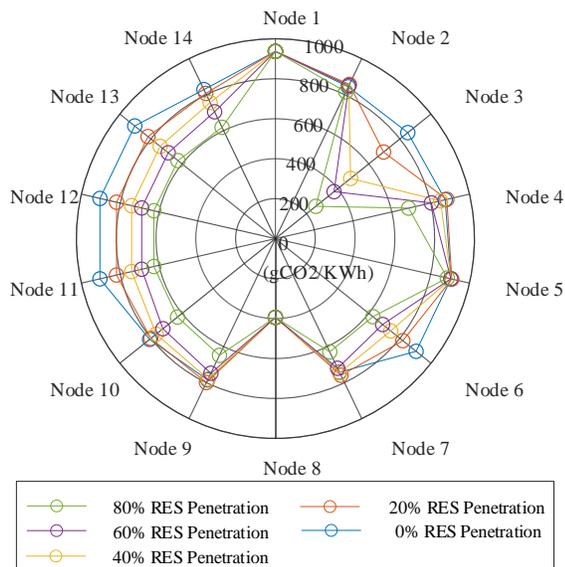

Fig 17. The average carbon intensity in different nodes.

### D. Carbon Footprint of EV Charging Behaviors

For the EV charging at Bus 9, 10, and 11, the system carbon emissions caused by charging stations under different scenarios are shown in Fig 18. Based on the analysis result, the carbon emissions caused by EVs also illustrated the non-linearity with system RES penetration. The carbon emissions from EVs decrease as the penetration of the RES system increases for charging stations located at Bus 10, and 11; however, the correlation between bus charging stations and the penetration of the RES is not complete as the emission caused by charging stations located at Bus 9 illustrated.

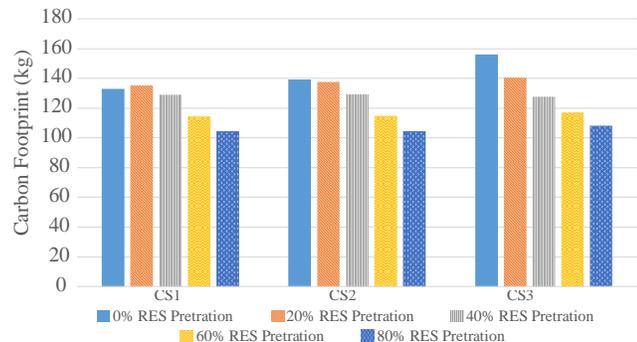

Fig 18. The carbon footprint of EV charging behaviors.

Generally, for the charging stations located at different buses, the emission responsibility is not identical distribution. The effect of EV charging load on power systems has spatial characteristics that follow the power systems topology.

## V. CONCLUSION

This paper has proposed the two novel stochastic carbon emissions tracing methods, which provide the probabilistic analysis tool for load-side carbon footprint for power systems:

1) The proposed stochastic carbon footprint tracing method can reveal the complex non-linear relationship between the system's RES penetration and the load carbon footprint under the extensive range power system topology.

2) The proposed ultra-efficient stochastic carbon footprint tracing method using virtual-bus methods significantly improved the computational efficiency of the proposed model by simplifying the topology in the calculation of the carbon footprint analysis of the power system. This paper makes innovative use of the generator marginal carbon emission model in the carbon footprint analysis to capture the impact of RES uncertainty on traditional generator carbon emissions via output fluctuation.

3) The impact of various factors on system carbon emissions, including RES penetration, distribution network base load, and EV charging load, is demonstrated in the case study by the proposed model. The results of the analysis illustrated that; for the system level, the increase in RES penetration would reduce the total system carbon emissions; however, there is a highly non-linear relationship between load-side carbon footprint and system RES penetration.

4) Based on the analysis results of the case study, the mechanisms and incentives of a carbon market anchored to the carbon emissions of the total system as a reference indicator may not be fair to the different parts of the load side. The economic penalties or revenues associated with an increase or decrease in overall system carbon emissions should not be applied equally to all load-side components of the system. This finding provides a new perspective on the establishment of carbon market mechanisms and the implementation of incentives.